# Imaging cytochrome C oxidase and $F_oF_1$-ATP synthase in mitochondrial cristae of living human cells by FLIM and superresolution microscopy

Franziska Förtsch[a,b], Mykhailo Ilchenko[a], Thomas Heitkamp[a], Silke Noßmann[c], Birgit Hoffmann[b], Ilka Starke[a], Ralf Mrowka[c,d], Christoph Biskup[b,d], Michael Börsch[a,d,*]

[a]Jena University Hospital, Single-Molecule Microscopy Group, Nonnenplan 2 - 4, 07743 Jena,
[b]Jena University Hospital, Biomolecular Photonics Group, Nonnenplan 2 - 4, 07743 Jena,
[c]Jena University Hospital, Experimental Nephrology Group, Nonnenplan 2 - 4, 07743 Jena,
[d]Center for Medical Optics and Photonics (CeMOP) Jena, Germany

## ABSTRACT

Cytochrome C oxidase and $F_oF_1$-ATP synthase constitute complex IV and V, respectively, of the five membrane-bound enzymes in mitochondria comprising the respiratory chain. These enzymes are located in the inner mitochondrial membrane (IMM), which exhibits large invaginations called cristae. According to recent cryo-tomography, $F_oF_1$-ATP synthases are located predominantly at the rim of the cristae, while cytochrome C oxidases are likely distributed in planar membrane areas of the cristae. Previous FLIM measurements (K. Busch and coworkers) of complex II and III unravelled differences in the local environment of the membrane enzymes in the cristae. Here, we tagged complex IV and V with mNeonGreen and investigated their mitochondrial nano-environment by FLIM and superresolution microscopy in living human cells. Different lifetimes and anisotropy values were found and will be discussed.

## 1 INTRODUCTION

ATP (adenosine triphosphate) is the main energy supply for active processes in cells. The millimolar concentration of ATP is maintained predominantly by the membrane-bound enzyme $F_oF_1$-ATP synthase. Found in prokaryotes and eukaryotes, $F_oF_1$-ATP synthases are located in the plasma membrane of bacteria, in the thylakoid membrane of chloroplasts, or in the inner mitochondrial membrane (IMM). Four additional enzyme complexes (i.e. complex I to IV) are required to generate the so-called proton motive force (PMF) which drives the catalytic process of making ATP from ADP (adenosine diphosphate) and phosphate (Pi) by the mitochondrial $F_oF_1$-ATP synthase. The PMF comprises a proton concentration difference (ΔpH) across the IMM plus an electrical potential. The PMF is produced by a coupled series of redox processes through the enzyme complexes I to IV by the associated active proton pumping across the IMM. The entire process of making ATP by the collaborative work of these five membrane enzymes is called Oxidative Phosphorylation (OXPHOS).

Mitochondria are organelles within the eukaryotic cell. They adopt different shapes, i.e. change between tubular elongated species with diameter around 500 nm and several μm length and a smaller fragmented form of mitochondria (Fig. 1a). Two membranes with distinct composition and function confine the inner part of the mitochondrion which is called the mitochondrial matrix. While the outer membrane controls import and export of proteins and small molecules, the inner membrane contains the OXPHOS complexes and exhibits the PMF for ATP synthesis (Fig. 1b). The IMM is folded into cristae to enlarge the membrane surface area and to optimize ATP production. Imaging mitochondria and the ultrastructure of the cristae with nanometer resolution is achievable by electron microscopy, either by negative staining[1] (Fig. 1a) or by electron cryotomography[2].

* email: michael.boersch@med.uni-jena.de; http://www.single-molecule-microscopy.uniklinikum-jena.de

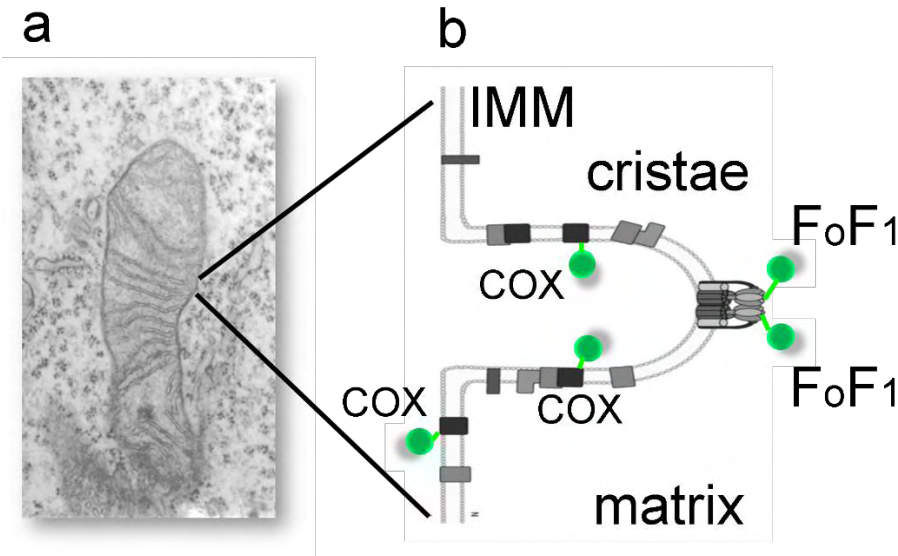

**Figure 1**: **a**, Electron microscopic image of a mitochondrion in a fixed human cell (HeLa cell, negative stain[1]). The diameter is approximately 500 nm. The cristae membranes are mostly oriented horizontally in this image. **b**, model of mitochondrial cristae in the IMM, with $F_oF_1$-ATP synthases located as dimers at the rim of the cristae[2] and cytochrome C oxidases in distinct planar areas of the IMM. Highlighted with green balls are fluorescent proteins mNeonGreen fused to either subunit γ of $F_oF_1$-ATP synthase or subunit 8a of cytochrome C oxidase in our study (modified from[3]).

Imaging mitochondria and the ultrastructure of the cristae with nanometer resolution is achievable by electron microscopy, either by negative staining (Fig. 1a) or by cryo electron tomography. Recently arrangement and localization of cytochrome C oxidase and $F_oF_1$-ATP synthase was visualized by electron cryotomography showing rows of dimeric $F_oF_1$-ATP synthases at the rim of the cristae[4], whereas complex I was found in planar areas of the cristae which are oriented perpendicular with respect to the outer membrane[2].

To monitor the distribution of fluorescently tagged OXPHOS complexes in the cristae of living cells, conventional optical microscopy is not sufficient due to the diffraction-limited resolution in the range of 200 nm[5-7]. Structured illumination microscopy[8] (SIM) or other superresolution microscopy techniques like STED[9] or PALM / STORM are necessary to resolve the stacking of the cristae. However, imaging cristae morphology has to be complemented by measuring additional fluorescence parameters for functional imaging[10, 11]. For example, measuring the distance between neighboring OXPHOS complexes can be achieved by fluorescence lifetime imaging[12] (FLIM) or anisotropy imaging based on Förster resonance energy transfer (FRET). Here we show that SIM can monitor reversible morphological changes of mitochondria induced by the addition of the antibiotic valinomycin, and FLIM as well as anisotropy imaging provides additional information on distinct biochemical states of individual mitochondria within the same human cell.

## 2 EXPERIMENTAL PROCEDURES

**2.1 Plasmid construction**

The DNA fragment encoding the γ-subunit of the mitochondrial $F_oF_1$-ATP synthase variant 2 was amplified by the polymerase chain reaction (PCR) using plasmid *pcDNA3.1-C-(k)DYK_ATP5C1V02* obtained from GenScript® as the respective DNA template. Using the restriction enzymes HindIII (New England Biolabs) and KpnI (New England Biolabs) the γ-subunit of $F_oF_1$-ATP synthase variant 2 was tagged on the C terminus with mNeonGreen using a linker sequence Gly-Thr-Thr-Arg-Pro-

Asp-Ile in between. The DNA sequence of the fluorescent protein mNeonGreen[13] was inserted after amplification of the mNeonGreen *pACWUBH1* vector[14, 15], kindly provided by Dr. Gabriele Deckers-Hebestreit (University of Osnabrück), with restriction endonuclease EcoRV (New England Biolabs) and AvaI (New England Biolabs) in the expression vector *pcDNA5/FRT* (Thermo Fisher Scientific), which was designed for use with *Flp-In$^{TM}$* system (Thermo Fisher Scientific). Correctness of sequences was confirmed by restriction analysis and by DNA sequencing. A similar mutant has been published previously[5].

The coding region of subunit 8a (Cox8a) from cytochrome C oxidase (complex IV) was amplified from cDNA of HeLa cells with the according primers and cloned into the expression vector *pcDNA5/FRT* (Thermo Fisher Scientific). The C terminus of subunit 8a was tagged with mNeonGreen using restriction endonucleases EcoNI (New England Biolabs) and AvaI (New England Biolabs). Correctness of sequences was confirmed by restriction analysis and by DNA sequencing. A similar mutant has been published previously[5].

**2.2 Stable transfection of HEK293-FRT cells by homologous recombination**

For stable transfection homologous recombination via the *Flp-In$^{TM}$* system (Thermo Fisher Scientific) was used. The *Flp-In$^{TM}$* system, which originally derives from yeast, provides the possibility to integrate a single and complete gene copy of a specific gene of interest (DNA sequence) into the genome and thus to generate a constitutive expression cell line. This cell line is isogenic and expresses the protein of interest stably. As a mammalian cell line, HEK293-FRT cells (Thermo Fisher Scientific) were used. Both the expression vector *pcDNA5/FRT* (Invitrogen/Thermo Scientific) and the HEK293-FRT cell line contain a Flippase Recognition Target site (FRT). The Flippase gene was provided by an additional vector pOG44 (Thermo Fisher Scientific) which was co-transfected. The enzyme recognizes FRTs and cuts the DNA. The expression vector containing the gene of interest is then inserted into the genome via Flippase recombinase–mediated DNA recombination at the FRT site. The resulting transgenic HEK293-FRT cells could be selected by hygromycin B due to the resistance gene cassette.

For the generation of stable expression clones, HEK293-FRT cells were seeded on 75 cm$^2$ cell culture flasks and grown in DMEM without phenol red (Gibco), with 10% FCS (Biochrom) and antibiotics (Penicillin/Streptomycin, Biochrom) till 50-90% confluence. For the preparation of the transfection solution 600 µl of DMEM (Gibco), 2 µg of the expression vector pcDNA5/FRT and 18 µg of pOG44 plasmid were mixed. In addition 600 µl of DMEM were added to 40 µl Roti®-Fect (Roth) and mixed. One day after both solutions were compiled, incubated for 20 min. at room temperature and dropwise added to HEK293-FRT cells, the transfection medium was replaced by fresh complete medium (DMEM, Gibco) and the cells were cultured for another day. After splitting and growing of the cells, 10 ml of complete medium with 300 µg/ml hygromycin B (Roth) were added to the cells. The medium was replaced again by 10 ml complete medium supplemented with 100 µg/ml hygromycin B after 24 hours.

**2.3 Cell culture**

HEK293-FRT cells were maintained in complete cell culture media Dulbecco's modified Eagle's medium (DMEM, Gibco), supplemented with 10% FCS (Biochrom) and 1% antibiotics (Penicillin/Streptomycin, Biochrom) and cultured at 37 °C. After 3 days cells were split and complete cell culture media was refreshed. Stable transfected HEK293-FRT were grown in complete cell culture media supplemented with 100 µg/ml hygromycin B (Roth). Transient transfection was performed by the common transfection method using calcium phosphate.

**2.4 Microscopy**

Structured illumination microscopy was performed on a Nikon N-SIM / N-STORM microscope[15-17] using 488 nm laser excitation and the recommended Nikon 60x water immersion objective with N.A. 1.27. The 1.5-fold magnification lens was combined with the 2.5-fold magnification lens for the 3D-SIM mode. Images were recorded by a cooled Andor EMCCD camera (iXon DU-897) using the Nikon FITC optical filter set. SIM images were recorded at 23.0±0.2°C. Nikon analysis software was used for SIM image reconstruction and for 3D visualization using the maximum intensity projection option.

Initial FLIM imaging was performed on a confocal LEICA TCS SP 8 microscope using a tunable pulsed supercontinuum laser.

For time-resolved anisotropy imaging, a custom-designed confocal microscope with 3D piezo scanning system (Physik Instrumente) mounted on an Olympus IX71 with 60x water immersion objective (n.a. 1.2) was used[18-22]. Ps-pulsed excitation was provided with 488 nm at 40 MHz (PicoTa 490, Picoquant). Fluorescence photons were detected by three single-photon counting avalanche photodiodes (SPCM-AQRH-14, Excelitas or Perkin-Elmer) for simultaneous spectrally-resolved FLIM and time-resolved anisotropy measurements[23]. Three synchronized TCSPC cards (SPC 153, Becker&Hickl) recorded the photons[24-26]. FLIM and time-resolved anisotropy images were analyzed using the SPCImage software (Becker&Hickl), whereas intensity-based anisotropy imaging was analyzed using counter cards (National Instruments) and custom Matlab scripts (Mathworks)[27-35].

## 3 RESULTS

To follow changes of the mitochondrial ultrastructure like variations in length or cristae remodelling *in vivo*, a bright and photostable fluorophore attached to the OXPHOS complexes can be used. As shown previously, fluorescent proteins subunit 8a of cytochrome C oxidase c can be fused genetically to the C terminus of subunit 8a of cytochrome C oxidase without impairing catalytic function[5]. Alternatively, the C-terminus of the γ-subunit of the $F_oF_1$-ATP synthase can be tagged with a fluorescent protein[36]. We have used a novel green fluorescent protein mNeonGreen[13] as a fluorophore attached to the $F_oF_1$-ATP synthase from *Escherichia coli* for single-molecule FRET studies of conformational changes *in vitro*[15, 37-41]. A higher extinction coefficient combined with a higher fluorescence quantum yield makes mNeonGreen a promising fluorophore for superresolution imaging in living cells.

### 3.1 Spectra, fluorescence lifetime and time-resolved anisotropy of soluble mNeonGreen

First we prepared a soluble protein variant of mNeonGreen using a Histag for protein purification. The photophysical properties of this soluble mNeonGreen in buffer are summarized in Fig. 2. At pH 7.5, the fluorescence excitation maximum was found at 506 nm and the fluorescence emission maximum was found at 518 nm (Fig. 2a). Steady-state fluorescence measurements were carried out on a modular fluorescence spectrometer (QuantaMaster 30, PTI). Fluorescence anisotropy excitation and emission spectra in cuvettes showed a constant anisotropy of r = 0.31 across the excitation and emission bands (Fig. 2b, c), as expected for a monomeric fluorescent protein with its internal rigid chromophore. Changing the pH of the buffer from 8.5 to 5.5 resulted in an increasing loss of brightness (or quantum yield) for a pH<7.0, but brightness remained constant for a pH above 7.0 (Fig. 2d, e).

The fluorescence lifetime of $\tau = 3.35 \pm 0.05$ ns (monoexponential decay fitting was appropriate) was measured in cuvettes using a pulsed LED at 503 nm with 10 MHz repetition rate in the spectrometer (FluoTime 200, Picoquant). Excitation pulse and fluorescence decay curve with fitting are shown in Fig. 2f. We tested a possible pH dependence of the fluorescence lifetime related to the brightness loss of mNeonGreen, but we found only a small increase from $\tau = 3.35$ ns at pH 8.5 to $\tau = 3.40$ ns at pH 5.5. Similarly, varying the NaCl concentration in the buffer (10 mM phosphate at pH 7.5) between 0 mM and 200 mM did not affect the fluorescence lifetime (data not shown). However, adding different amounts of glycerol resulted in a significant shortening of the fluorescence lifetime, i.e. from $\tau = 3.41$ ns at pH 8.0 in the absence of glycerol to $\tau = 2.975$ ns at 50% glycerol content to $\tau = 2.68$ ns at 90% glycerol content (data not shown).

We also measured the time-resolved fluorescence anisotropy decay of mNeonGreen at pH 7.5. As shown in Fig. 2g, soluble mNeonGreen exhibited an intrinsic anisotropy of $r_0 = 0.33 \pm 0.01$ (i.e. determined for time t = 0 ns set at the maximum of the excitation pulse). The slow anisotropy decay corresponded to a rotational correlation time $\tau_{rot} = 14.8 \pm 2.6$ ns, in good agreement for similar sized, soluble green fluorescent proteins[42]. In contrast, a rotational correlation time $\tau_{rot} = 48 \pm 7$ ns was measured for mNeonGreen fused to the purified solubilized $F_oF_1$-ATP synthase from *E. coli* in buffer containing the detergent DDM. For this mNeonGreen-tagged $F_oF_1$-ATP synthase with a twenty times larger molecular weight compared to the soluble protein mNeonGreen, the intrinsic anisotropy was $r_0 = 0.32 \pm 0.08$ (data not shown).

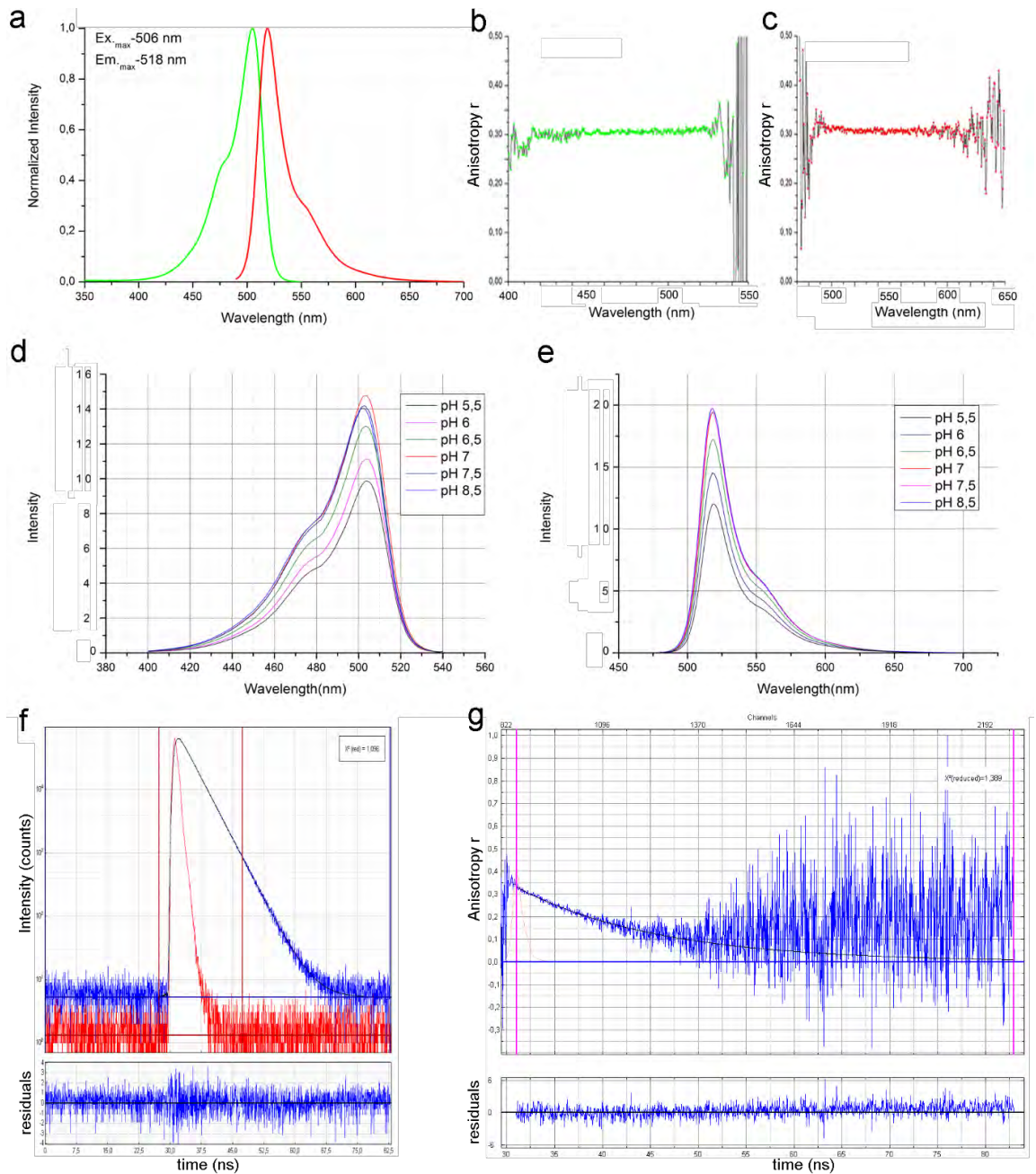

**Figure 2**: **a**, normalized fluorescence excitation (green) and emission (red) spectra of soluble mNeonGreen in buffer. **b**, fluorescence excitation anisotropy spectrum, and **c**, fluorescence emission anisotropy spectrum o mNeonGreen. **d**, pH-dependent brightness changes in the fluorescence excitation spectra and **e**, in the fluorescence emission spectra. **f**, fluorescence lifetime decay (blue) with fit (black) of soluble mNeonGreen. Residuals are shown in the lower panel, IRF of the 503 nm pulsed LED in red. **g**, time-resolved anisotropy decay of soluble mNeonGreen (blue) with fit (black) and residuals in the lower panel.

## 3.2 FLIM imaging of mNeonGreen fused to cytochrome C oxidase or $F_oF_1$-ATP synthase in mitochondria and of soluble mNeonGreen in the cytosol of HEK293 cells

Stably transfected HEK293 cells either expressing mNeonGreen fused to cytochrome C oxidase subunit 8a or to the γ-subunit of $F_oF_1$-ATP synthase were generated using the *Flp-In*$^{TM}$ system. Therefore, each cell contained the targeted mitochondrial membrane protein and we did not need to search for an appropriate cell using the confocal microscope. As seen in Fig. 3a and b, both constructs for mNeonGreen fused to $F_oF_1$-ATP synthase or to cytochrome C oxidase, respectively, exhibited strong fluorescence signals in the mitochondria of the cells. The averaging process for accumulated FLIM measurements over several minutes resulted in blurred images of the mitochondria due to the motion of the organelles within the living cells. Fluorescence lifetimes had to be fitted by double-exponential decays, resulting in average lifetimes of $\tau_{(avg)}$ = 2.8 ns for both mutated enzymes. The longer lifetime component $\tau$ = 2.9 ns was dominating with amplitudes of about 90%. The shorter lifetime component was about $\tau$ = 1.3 ns to 1.6 ns. We also expressed soluble mNeonGreen transiently in the cytosol of HEK293 cells and determined fluorescence lifetimes $\tau_{(avg)}$ = 2.93 ns, with the longer component $\tau$ = 3.04 ns (~ 80%) and the shorter component $\tau$ = 1.4 ns (Fig. 3c).

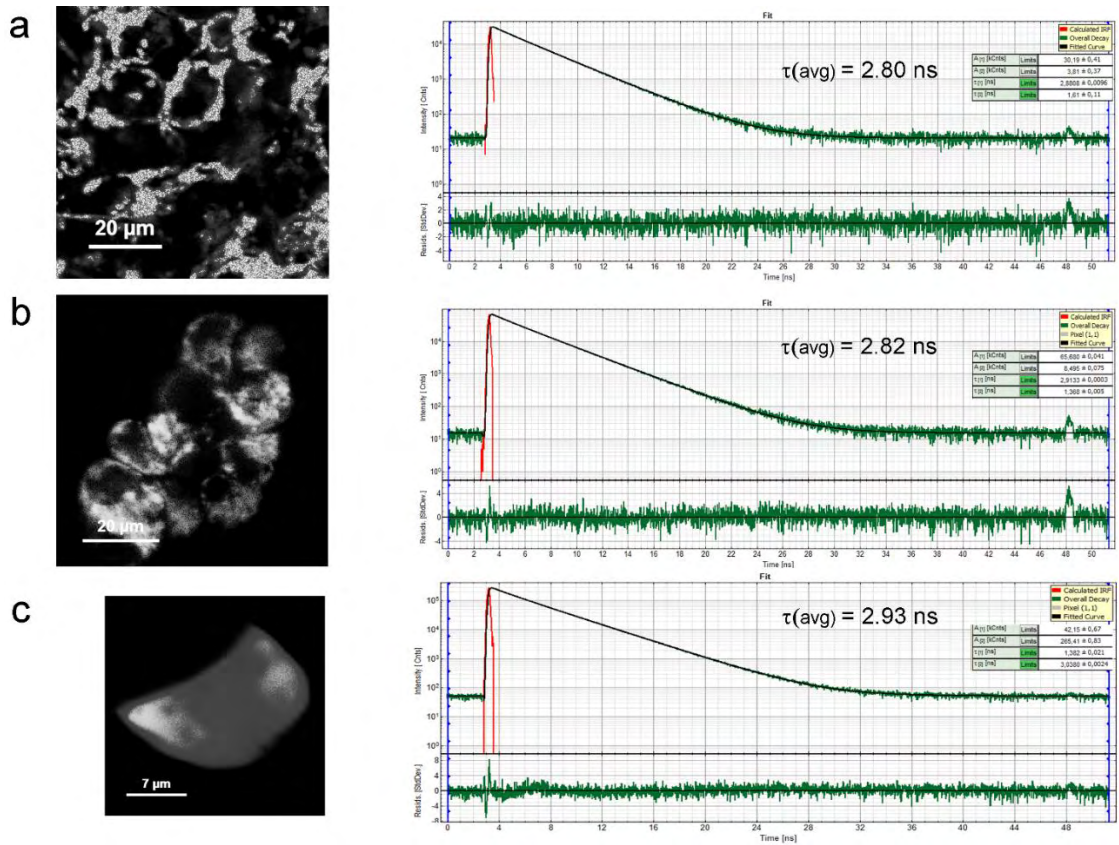

**Figure 3**: **a**, FLIM image and associated lifetime decay (all highlighted pixels, with double-exponential fit) of mNeonGreen fused to the γ-subunit of $F_oF_1$-ATP synthase in stably transfected HEK293 cells at room temperature. **b**, FLIM image and associated lifetime decay (all highlighted pixels, with double-exponential fit) of mNeonGreen fused to subunit 8a of cytochrome C oxidase in stably transfected HEK293 cells at room temperature. **c**, FLIM image and associated lifetime decay (all highlighted pixels, with double-exponential fit) of soluble mNeonGreen in the cytosol of transiently transfected HEK293 cells at room temperature.

## 3.3 SIM Superresolution imaging of mNeonGreen fused to $F_oF_1$-ATP synthase in mitochondria

The cristae of mitochondria are the invaginated parts of the inner membrane and accommodate the $F_oF_1$-ATP synthases and the cytochrome C oxidases. In electron microscopic images of mitochondria in human cells the cristae appear to be oriented perpendicular to the outer membrane (Fig. 1a). Accordingly, labeled $F_oF_1$-ATP synthases or cytochrome C oxidases are expected to result in a stripe-like intensity pattern of mitochondria along the long axis. Because of the short distance between neighboring cristae, superresolution microscopy like SIM, STED or STORM / PALM is required to reveal these stripes.

We applied SIM to look for stripe-like intensity pattern in mitochondria marked by $F_oF_1$-ATP synthases or cytochrome C oxidases. Using the 3D-SIM mode of the Nikon N-SIM microscope, 15 individual sub-images with 100 ms acquisition time had to be recorded in order to obtain one reconstructed SIM image in one image plane. A z-stack of 19 layers (or image planes, respectively) was collected to generate the height-encoded maximum intensity projection images representing the threedimensional volume image. About 2 min measurement time was necessary for a complete z-stack. Within this time some of the mitochondria moved quickly in three dimensions but others stayed in place. A high laser power at 488 nm (with a 35% power setting in the Nikon software) was applied for the widefield illumination in SIM. However, repeated recording of the same cell was possible for up to 1 h despite increasing loss of fluorescence intensity.

Elongated mitochondria were found in living HEK293 cells using mNeonGreen-tagged $F_oF_1$-ATP synthases (Fig. 4). The relative height of the mitochondria within the cell is color-coded. Mitochondria near the plasma membrane were colored blue (0 to 1 µm above the cover glass), those in a distance range of 2 to 3.5 µm were shown in green, and mitochondria on the upper parts of the cell in the range of 4.5 to 5.7 µm were colored orange-red. The average diameter of the mitochondria was found in the range of 400 to 500 nm.

Comparing Figs. 4a and b, some of the mitochondria obviously moved within this cell changing their shape and the lateral position with shifts larger than 1 µm. We did not detect a stripe-like intensity pattern in these mitochondria due to motion blurring. In addition, the mitochondria were tightly packed which resulted in a large fluorescence contribution from out-of-focus mitochondria. This out-of-focus blur prevented a good SIM image reconstruction but allowed deconvolution of the SIM images only.

Valinomycin (10 µM) was added carefully as a droplet to the surface of the buffer medium covering the cell to allow for slow mixing and diffusion-controlled uptake by the cell. Immediately after addition of this $K^+$ ionophore, which is used for depolarization of electric membrane potentials in all cellular membranes including the IMM[43], no obvious changes in the shape of the mitochondria were found (Fig. 4c). However, after 5 min all mitochondria appeared round-shaped (Fig. 4e). They maintained the distorted morphology for a few minutes (Fig. 4f). After about 10 min (Fig. 4g), re-shaping of the mitochondria was observed. 15 min after addition of valinomycin, mitochondria appeared elongated again (Fig. 4i), and maintained that shape until the end of our SIM image recording (1 h).

Using mNeonGreen fused to cytochrome C oxidase we observed the same behavior upon addition of valinomycin to the HEK293 cells. Before addition, the mitochondria appeared elongated, but change to round-shaped forms with brightest parts on the rim within a few minutes. They recovered an elongated shape after about 10 minutes.

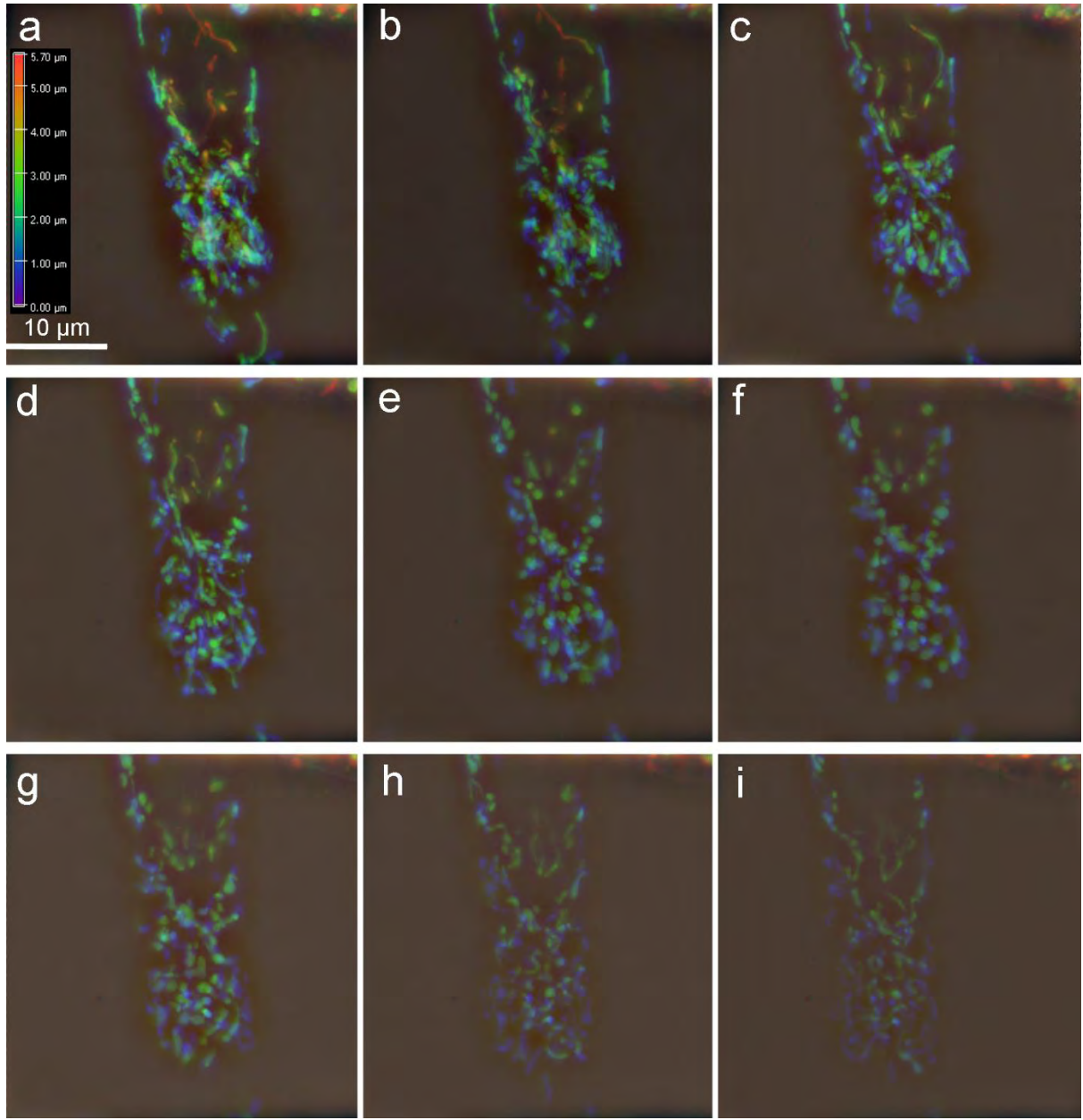

**Figure 4**: Time series of superresolution images (SIM) showing mitochondria in a HEK293 cell. mNeonGreen was fused to the γ-subunit of $F_oF_1$-ATP synthase. Each image comprises a z-stack with 19 layers, all 300 nm apart. The false color is coding the height information, with blue color mitochondria located near the cover glass up to red ones that were 5.7 µm above the cover glass. Shown is the maximum intensity projection of the threedimensional volume image. **a**, 10 min; **b**, 5 min before, and **c**, immediately after addition of 10 µm valinomycin. **d**, 3 min, **e**, 5 min, **f**, 7 min, **g**, 9 min, **h**, 12 min, and **i**, 15 min after addition of valinomycin.

## 3.4 FLIM and anisotropy imaging of mNeonGreen fused to cytochrome C oxidase in mitochondria

While SIM allowed to image the varying shapes and movements of mitochondria in living cells, additional information about the physiological state of the mitochondria is needed. Therefore we measured the fluorescence lifetime of our mNeonGreen-tagged membrane proteins in the IMM. Our custom-designed confocal microscope used a pulsed, polarized 488 nm diode laser for ps pulses at 40 MHz, coupled into an Olympus IX microscope that was equipped with a 60x water immersion objective (N.A. 1.2, Olympus). Here we were using only two out of four APDs of the setup to record the photons in the range of 500 to 570 nm and to separate them by a polarizing beam splitter for FLIM measurements and time-resolved anisotropy recordings. This instrument is mostly used for single-molecule FRET measurements *in vitro* [34, 44-57]. A piezo scanner (100x100 µm scan range) and a piezo-driven objective positioner allowed to record the FLIM and anisotropy images in parallel at arbitrary pixel size as well as z-stacks for three-dimensional volume imaging.

Fig. 5 shows the time-dependent changes of mitochondrial positions and shapes following addition of 10 µM valinomycin.

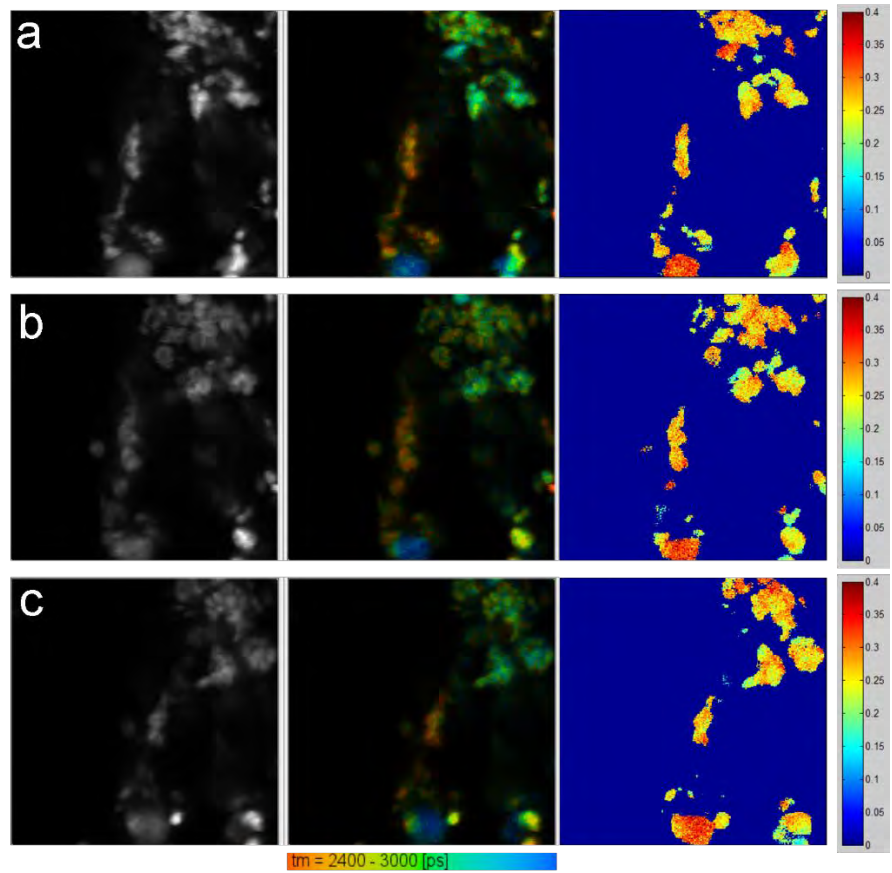

**Figure 5**: Time series of FLIM and steady-state anisotropy images showing mitochondria in a HEK293 cell. mNeonGreen was fused to subunit 8a of cytochrome C oxidase. Left panels are intensity images (grey). The false color is coding for the mean fluorescence lifetime (middle panels) or the intensity-based fluorescence anisotropy per pixel (right panels). Image size is 25x25 µm$^2$. Conditions: **a**, 5 min before, **b**, immediately after addition of 10 µm valinomycin, and **c**, 10 min after addition of 10 µM valinomycin.

Before addition of valinomycin (Fig. 5a), some mitochondria were clearly elongated, especially in cellular areas without too many neighboring mitochondria. In the upper right part of the image, the mitochondria were crowded. Here, motion blur of moving mitochondria was significant because we used the slow piezo scanner with 4 ms dwell per pixel.

The fluorescence lifetime was different in distinct mitochondria, but appeared similar within a single mitochondrion (medium panels, false-colored). The elongated mitochondria in the middle of the image exhibited shorter mean lifetimes around 2.4 ns and were lifetime-colored orange-red. In contrast, the mitochondria in the upper left (green colored) showed longer mean lifetimes around 2.7 ns, whereas only one large round-shaped organelle exhibited a 3 ns lifetime (lower part of the image in Fig. 5a).

After addition of 10 µm valinomycin (Fig. 5b), the mitochondria were found round-shaped, with a brighter rim as seen in the upper part of the intensity (left) and the FLIM (middle) images. The mean fluorescence lifetime of the individual mitochondria did not change in comparison to the FLIM image before valinomycin addition. For example, those mitochondria with a shorter lifetime maintained this lifetime throughout the massive rearrangement of the cristae of the inner mitochondrial membrane. Due to the long scanning time for each image in the range of nearly 5 minutes, morphological changes were identified more clearly at apparently shorter time scales after addition of valinomycin when compared to the time-dependent SIM image series in Fig. 4. About 10 minutes after addition of valinomycin, mitochondria were shrinking and some recovered elongated shapes (Fig. 5c). The distinct fluorescence lifetimes did not change with the re-shaping process of the mitochondria, as seen for the red-colored mitochondria in the middle of the image exhibiting an elongated shape again.

Simultaneous to the FLIM image the fluorescence anisotropy image was measured using the intensity per pixel information. Polarized emission was recorded by the two APDs and an apparent steady-state anisotropy $r$ was calculated for each pixel using

$$r = \frac{I(parallel) - I(perpendicular)}{I(parallel) + 2*I(perpendicular)} \quad (1)$$

without further corrections, i.e. without correction for a measured G factor for the slightly different detection efficiencies for fluorescence intensities polarized parallel to the polarization of the exciting laser [*I(parallel)*] or perpendicular [*I(perpendicular)*], and, importantly, without corrections for the high numerical aperture of the microscope objective.

The intensity-based anisotropy images are shown as the right panels in Fig 5. Most mitochondria exhibited high anisotropies within the range of $r \sim 0.3$ (pixels colored red). However, many mitochondria showed lower anisotropy values in the range of $r \sim 0.2$ (green) to 0.25 (yellow), i.e. lower than expected for mNeonGreen. Very few mitochondria exhibited anisotropy values around $r \sim 0.15$ (cyan). The sequential reshaping processes of the mitochondria induced by valinomycin addition did not change the fluorescence anisotropy for individual mitochondria, at least no drastic changes could be identified.

# 4 DISCUSSION

Mitochondria are the powerhouse of the cell because they produce most of the "energy currency" ATP. The OXPHOS enzymes I to V are located in the cristae of the inner mitochondrial membrane and and work together to build-up the PMF which drives ATP synthesis by $F_oF_1$-ATP synthases (i.e. complex V). Tagging cytochrome C oxidase or $F_oF_1$-ATP synthases with mNeonGreen allowed us to image the mitochondria in living HEK293 cells. The brightness of mNeonGreen was high enough for widefield laser excitation required for superresolution imaging by SIM as well as for confocal pulsed excitation for FLIM and (time-resolved) fluorescence anisotropy imaging. We monitored mitochrondrial movements in the cells for up to 1 hour.

SIM imaging revealed the elongated shape and the dynamic re-arrangement of mitochondria within the cell. In starving cells, stripe-like intensity pattern were found in some mitochondria of HEK293 cells indicating that cristae were eventually resolvable with superresolution microscopy of mNeonGreen-tagged mitochondrial membrane proteins. When imaging mitochondria in the presence of growth medium we rarely resolved these stripe-like patterns. Because recording one SIM image layer comprising 15 sub-images with the shifted and rotated grating took about 2 seconds, a z-stack of 19 layers to cover 5.7 µm height was completed in about 2 minutes. During this time span mitochondria moved in all spatial dimensions. Reconstructing the SIM mitochondria images was limited and disturbed by the densely packed arrangement of the mitochondria that was causing strong out-of-focus fluorescence. In future work we will switch from HEK293 cells to very flat cell lines exhibiting only one layer of mitochondria in a given area of the cell.

We added the ionophore valinomycin at a high concentration of 10 µm to the HEK293 cells to trigger a morphological change of the mitochondria. At concentrations below 100 nM, valinomycin destroys the $K^+$ gradient across all cellular membranes including the inner mitochondrial membrane, and, as a consequence, swelling of mitchrondria has been reported[43]. Here we found that swelling can be monitored in real time by SIM. The breakdown of the $K^+$ concentration difference across the IMM resulted in round-shaped mitochondria with bright fluorescence located at the rim. However, within 10 to 15 min the cells apparently restored the primary shape and maintained elongated forms of mitochondria for up to one hour until the end of image recording.

Confocal FLIM imaging confirmed the valinomycin-induced reversible change of mitochondrial shape. Fluorescence lifetimes were fitted using a double exponential decay function for both the FLIM images of the fast scanning Leica microscope (using SymPhoTime 64 software, Picoquant) or the SPCImage software (Becker&Hickl) in our custom-designed piezo scanning microscope. Excitation with 488 nm should preferentially excite the mNeonGreen label on cytochrome C oxidase of $F_oF_1$-ATP synthase. However, partial excitation of endogenous fluorescent redox cofactors like $FAD^+$ or NAD(P)H in mitochondria[58, 59] cannot be excluded. These cofactors might explain the short lifetime component of 1.4 to 1.6 ns as seen in Fig. 3. The mean lifetime was different for individual mitochondria the HEK293 cells so that they were trackable in the cell throughout the FLIM recording. Mitochondrial reshaping to round forms after valinomycin addition and back to elongated forms did not result in significant changes of the fluorescent lifetimes, indicating that the local membrane environment of the tagged enzymes was maintained during the enormous morphological changes.

Steady-state anisotropy images revealed the expected high anisotropy values for membrane protein-bound fluorophores in the mitochondria. However, a few mitochondria showed lower anisotropy values around r ~ 0.15. Presently we have no explanation for this observation. We started to analyze the time-resolved anisotropy decays using photons collected from several mitochondria. Thereby we retrieved smooth decay curves required for the calculation of time-resolved anisotropy data. Binning a few mitochondria with apparently the same mean fluorescence lifetime and similar steady-state anisotropy was sufficient to find at least two distinct populations of mitochondria with different time-resolved anisotropy behavior. Analyzing the local nano-environment by time-resolved anisotropy imaging might allow for *in vivo* detection of OXPHOS supercomplexes[60, 61] in the cristae and their rearrangement upon chemically induced stress or mitochondrial aging processes in general.

**Acknowledgement**

The authors thank all members of our research groups who participated in various aspects of this work, from genetics and biochemistry support to cell culture and maintenance of the microscopes. Financial support for the Nikon N-SIM / N-STORM by the State of Thuringia (grant FKZ 12026-515 to M.B.) is gratefully acknowledged.